\documentclass[conference]{IEEEtran}
\IEEEoverridecommandlockouts
\usepackage{multirow}
\usepackage{cite}
\usepackage{amsmath,amssymb,amsfonts}
\usepackage{algorithmic}
\usepackage{graphicx}
\usepackage{textcomp}
\usepackage{xcolor}
\def\BibTeX{{\rm B\kern-.05em{\sc i\kern-.025em b}\kern-.08em
    T\kern-.1667em\lower.7ex\hbox{E}\kern-.125emX}}
\begin{document}

\title{Jamming Pattern Recognition over  Multi-Channel Networks: A Deep Learning Approach \\
\thanks{This research was supported in part by the U.S. Office of Naval Research  under MURI Grant N00014-19-1-2621.}
}
\author{Ali Pourranjbar$^{1}$,Georges Kaddoum$^{1}$and Walid Saad$^{2}$\\
 
         \\$^{1}$Department of Electrical Engineering
ETS - University of Quebec, Montreal QC, Canada,\\
Emails: ali.pourranjbar.1@ens.etsmtl.ca, georges.kaddoum@etsmtl.ca.
             \\$^{2}$Wireless@VT, Bradley Department of Electrical and Computer Engineering,\\ Virginia Tech, Blacksburg, VA, USA, Email: walids@vt.edu.
}

\maketitle

\begin{abstract}
 With the advent of intelligent jammers, jamming attacks have become a more severe threat to the performance of wireless systems. An intelligent jammer is able to change its policy to minimize the probability of being traced by legitimate nodes. Thus, an anti-jamming mechanism capable of constantly adjusting to the jamming policy is required to combat such a jammer. Remarkably, existing anti-jamming methods are not applicable here because they mainly focus on mitigating jamming attacks with an invariant jamming policy, and they rarely consider an intelligent jammer as an adversary. Therefore, in this paper, to employ a jamming type recognition technique working alongside an anti-jamming technique is proposed. The proposed recognition method employs a recurrent neural network  that takes the jammer's occupied channels as inputs and outputs the jammer type. Under this scheme, the real-time jammer policy is first identified, and, then, the most appropriate countermeasure is chosen. Consequently, any changes to the jammer policy can be instantly detected with the proposed recognition technique allowing for a rapid switch to a new anti-jamming method fitted to the new jamming policy. To evaluate the performance of the proposed recognition method, the accuracy of the detection is derived as a function of the jammer policy
switching time.  Simulation results show the detection accuracy for all the considered users’ numbers is greater than $70\%$ when the jammer switches its policy every $5$ time slots and the accuracy raises to $90\%$  when the 
jammer’s policy switching time is $45$.
\end{abstract}


\section{Introduction}
Wireless communication networks are prone to jamming assaults due to their shared and open nature. Jammers degrade the performance of wireless communication networks in terms of transmission rate, secrecy rate, and throughput by transmitting disruptive signals to the utilized spectrum of legitimate nodes. As a result,   it is necessary to adopt an anti-jamming technique.  
Many effective anti-jamming methods, such as those in \cite{tcomkhodam, xuan2011trigger, d2014defeating, elleuch2021novel,c6} have been proposed. However, existing methods mainly focus on mitigating jamming attacks with an invariant jamming policy. For instance, the authors in  \cite{tcomkhodam, xuan2011trigger, d2014defeating} propose anti-jamming methods against reactive jammers. Other approaches, such as those in \cite{elleuch2021novel}  and \cite{c6}  that propose machine learning-based anti-jamming techniques against multi-type jammers, assume that jammers' policies are not changed during the interaction between legitimate users and jammers, and as a result, when the jamming policy changes, legitimate nodes need to be retrained. These anti-jamming techniques are not effective against jammers with dynamic jamming policies that continuously change their jamming policies to avoid being tracked. Therefore, it is essential to constantly monitor jammers' behavior and  adopt an anti-jamming method based on  their current jamming policies. This way, a jamming recognition approach is required  to identify the jammers’ policy.

Jamming recognition techniques are primarily studied in the context of radar \cite{shao2020convolutional ,wu2017jamming, wang2019recognition, qu2020jrnet}, focusing on the jammer types detection based on the jamming signals. The work in \cite{shao2020convolutional}  and \cite{wu2017jamming} study the jamming signal type recognition using convolution neural networks (CNN) while the authors in \cite{wang2019recognition} employ  CNN to classify the jamming signal based on  fast Fourier transform (FFT). The authors in \cite{qu2020jrnet} study power spectral of the jamming signals to recognize the jamming type. The proposed techniques in \cite{shao2020convolutional ,wu2017jamming, wang2019recognition, qu2020jrnet} need a high sampling rate with high accuracy of the signal detection. As a result,  legitimate users are required to detect the jamming signals with high accuracy, which is not possible in all scenarios, especially in cases where the jammers jam channels with low jamming power.  Thus, in this paper, we mainly focus on recognizing the patterns followed by jammers in selecting the frequency channels. Consequently, the legitimate nodes only need to detect the presence of the jammers in their utilized spectrum instead of detecting the jamming signals.

The authors of \cite{cai2019jamming} employ a CNN to predict a jamming pattern from  jammer's spectrum waterfall plots. Due to the robustness of CNNs in identifying patterns in images, the authors of \cite{cai2019jamming} utilize a spectrum waterfall plot to translate the jammer's occupied frequency channels to pictures. In spite of the fact that the authors of \cite{cai2019jamming} propose a  jamming recognition technique, they just  consider   three simple jamming scenarios, namely, sweeping, single tone, and multi-tone jamming. Moreover, the recognition technique proposed in  \cite{cai2019jamming} needs a significant dataset related to the behavior of the jammer. 

The main contribution of this paper is to develop a jamming recognition technique that can work with a small dataset, and that can extend to different jamming types.  To this end, we employ recurrent neural networks (RNNs) as they  best model the data with sequential nature in many cases. We then proposed an RNN based recognition technique that can recognize the jamming policy in a short period of time and easily extendable for any type of jamming policy.  We consider a system model composed of an  access point (AP) serving users in the presence of a jammer that constantly changes its jamming policy. We perform different simulations considering different jamming policies and a wide range of the jammer's policy switching times. Our results show that the proposed recognition method is able to properly find all of the considered jamming policies with high
accuracy within a short period.  
 
The rest of  the paper is organized as follows.  Sections II and III
 present  the system model and proposed jamming recognition  technique, respectively. Simulation results are
presented in Section IV, and finally, conclusions are drawn in
Section V.
 
\section{System Model}

We consider wireless networking consists of  AP, $N$  users, and a jammer, which are  uniformly distributed in space. We assume that the users will always have a packet to transmit and transmit, and they are able to  communicate with the AP over $L$ orthogonal frequency channels. Time is divided into equal time slots, and in each time slot, every user selects a frequency channel. Moreover, users hop between frequency channels and  they do not collide with each other. We consider a jammer with a dynamic jamming policy as the adversary. The considered jammer switches its jamming policy every $K$ time slots and selects a policy among considered policies randomly. We consider five different jamming policies:
\begin{itemize}
\item \emph{Random jammer (RJ)} that selects its channels randomly among available channels every time slot.
\item \emph{Sweeping jammer (SJ)} that shifts its full energy between multiple frequency channels in a cyclic manner.
\item \emph{Fast reactive jammer (FRJ)} that continuously listens to channels and jams channels immediately after sensing an activity.
\item \emph{Reactive jammer with delay (RJWD)} that behaves similar to the FRJ; however, it jams the detected channels after a single time slot from sensing an activity.
\item\emph{Combat jammer  (CJ)} that  selects a specified number of channels randomly and keeps jamming those channels for $M$ consecutive time slots.
\end{itemize}

\section{Proposed jamming recognition technique }
 In order to select an effective anti-jamming technique against a jammer with a dynamic  policy, we propose a jamming type recognition technique by monitoring the  frequency channels occupied by the users, and the jammer. The proposed recognition technique can be divided into two parts of training and testing. The training part is an offline process in which the AP simulates the interaction between itself, users, and the jammer. On the other hand,  testing is performed online during the interaction between the AP, users and the jammer. In the training process, the AP simulates the interaction between itself, users and the jammer.
 In order to model this interaction, we propose to use RNNs that are known to be suitable for analyzing data with sequential nature, as the case in our setting.
Particularly, RNNs are a class of
artificial neural networks capable of learning patterns and long-term dependencies from time series and sequential data. At time step $t$, the RNN gets previous  hidden state $\boldsymbol{h}_{t-1}$ and the input $\boldsymbol{x}_t$. Then, it produces   the updated hidden state $\boldsymbol{h}_{t }$  and output $\boldsymbol{y}_t$ as follows\cite{RNN}:
\vspace{-0.7cm}

\begin{equation}
\begin{aligned}
\\&\boldsymbol{h}_t =f(\boldsymbol{W}_h \boldsymbol{h}_{t-1}+\boldsymbol{V}_h\boldsymbol{x}_t +b_n)
\\&\boldsymbol{y}_t =f(\boldsymbol{W}_yh_t +b_y),
\end{aligned}
\end{equation}
where $f$ is an activation function, $\boldsymbol{V}_h$ is the weight of the input vector, $\boldsymbol{W}_h$ and  $\boldsymbol{W}_y$ are weights of the hidden layer and the output, respectively. Moreover, $\boldsymbol{b}_n$ and  $\boldsymbol{b}_y$ are both the bias terms.

\begin{figure}[!tbp]
\centering
 \includegraphics[width=0.5\textwidth, height=0.35\textwidth]{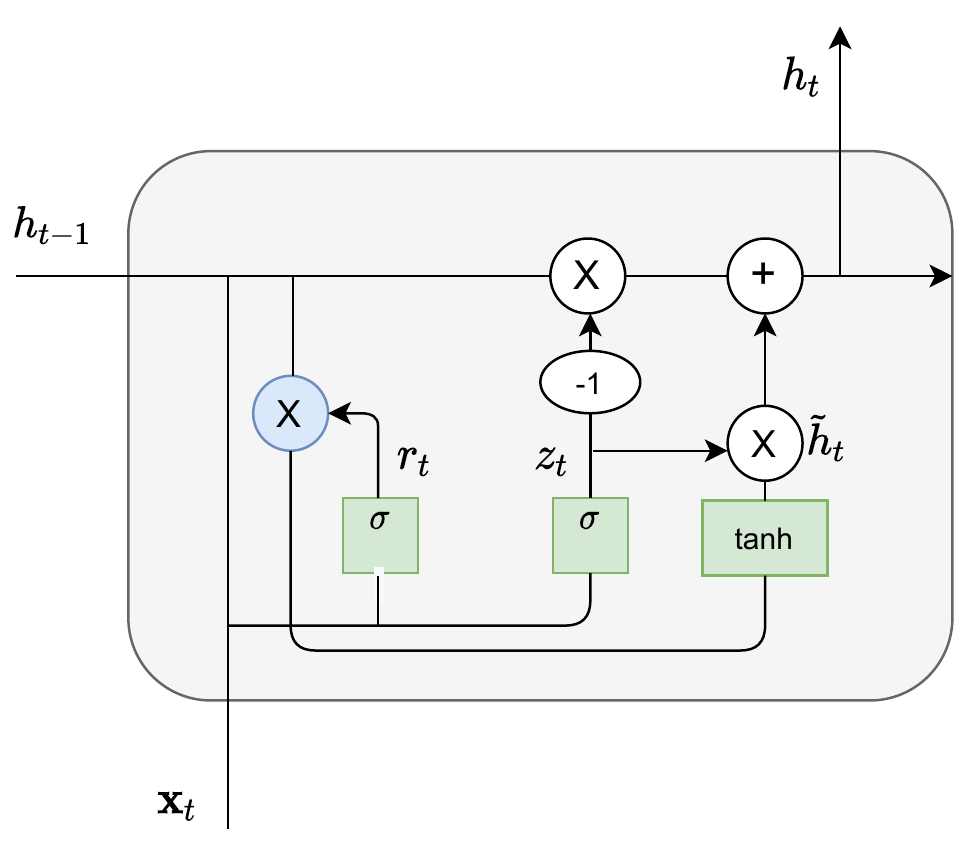}
\caption{ GRU cell unit. }
\label{fig2}
  \end{figure}

Vanishing gradients  is a major problem of conventional RNNs due to their high depth and recurrent connections. Long
short term memory (LSTM) and gated recurrent unit (GRU)
models mitigate the above-mentioned problem of vanishing
gradients by controlling the inputs using several gates in a
hidden layer  as depicted in Fig. \ref{fig2}. In this work, we employ GRU as the recurrent
unit due to the simplicity of its internal structure in comparison to
LSTM model, which results in faster training.  
The GRU architecture addresses this problem by adding    
a controller gate $z_t$ meaning that when $z_t$  is  $1$,  the input gate is closed, while  $z_t$ is $0$,  the input gate is opened. As a result, the relationship  between the input and output is changed to:
\vspace{-0.5cm}
\begin{equation}
\begin{aligned}
\\&r_t =\sigma(W_{xr}^T.\boldsymbol{x}_t +W^T_{hr}.h_{t-1}+b_r)\\&
z_t = \sigma(W_{xz}^T.\boldsymbol{x}_t +W^T_{z}.h_{t-1}+b_z) 
\\&\tilde{h_t} = \tanh(W^T_{x\tilde{h}}.\boldsymbol{x}_t +W^T_{h\tilde{h}}.(r_t  \otimes h_{t-1})+b_{\tilde{h}})
\\&h_t = z_t \otimes h_{t-1} +(1-z_t) \otimes \tilde{h_t},
\end{aligned}
\end{equation}
where $W_x r$, $W_xz$, $W_{xh}$, $W_hr$, $W_hz$,  and $W_h$  are learned weight matrices, $\delta$ is the Sigmoid function, and $b_r$ , $b_z$, $b$
are bias\cite{RNN}.

\begin{figure}[!t ]
\centering
 \includegraphics[width=0.4\textwidth, height=0.3\textwidth]{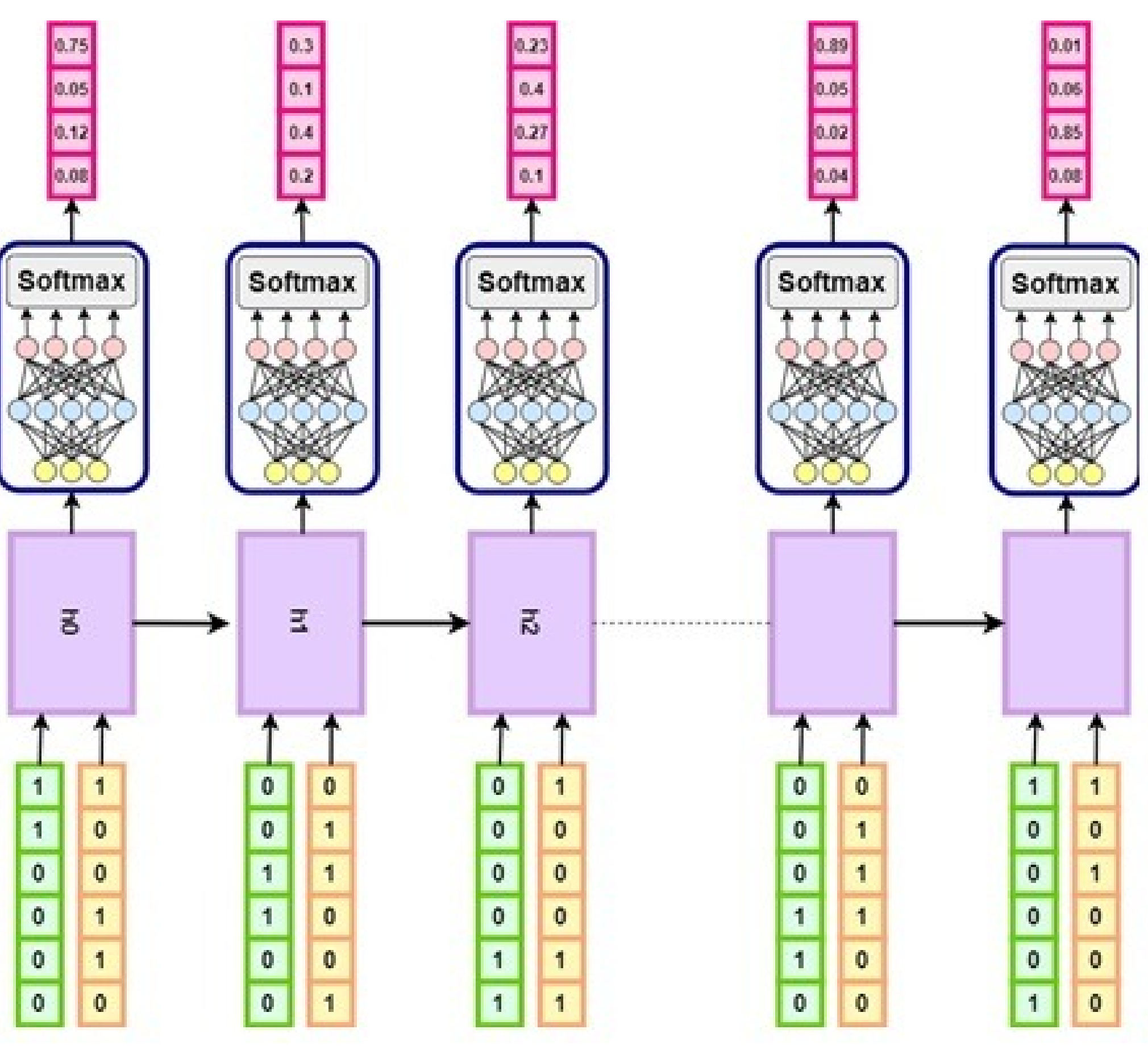}
\caption{Proposed RNN network. }
\label{fig3}
  \end{figure}

In each recurrent step, followed by the GRU hidden states, we apply a feed-forward layer with the output size equal to the number of considered jamming policies. Next, the output of the fully connected layer is fed into a SoftMax layer to  produce a probability distribution over the jamming type classes. It is noteworthy that the predicted class is expressed as
an index of the output array that has the highest probability value. The proposed network is depicted in Fig. \ref{fig3}.

As mentioned, the AP simulates the interaction between itself, the users, and the jammer for $T$ consecutive time slots. In this scheme, users select their channels randomly, and then the jammer reacts to the users' behavior according to its policy. The random policy is selected for the users' channel allocation to generalize the model. The AP saves the occupied channels of the users, the jammer, and the corresponding jamming policy during the simulation. In more detail, the AP considers a  vector with $2L+1$ zero elements in each time slot.
Then, in the first $L$ elements, it stores/saves the users' utilized channels by setting the elements corresponding to occupied channel indices to one, and in the following $L$ element, it repeats the procedure for the jammer's occupied channels.
Moreover, in the element with an index of $2L+1$, the AP saves the jamming policy class, a number between one and five.

After  the simulation is done, the collected data is merged together vertically and    utilized to train the proposed RNN. In this context, in order to   train  the RNN, all the elements with column indices of $1$ to $2L$, and  elements with  column  index $2L+1$ are utilized as the input and the target of the proposed network, respectively. In each step of training, a $P$ number of consecutive vectors are fed to the RNN and  $P$ vectors with five elements are outputted from the network. The cross-entropy is utilized as the loss function to train the network.

After that, the proposed network is trained, and the trained network is employed to recognize the jamming types. In the testing process,  similar to the training process, the AP saves users and the jammer behavior for $C$ last time slot. In each time slot of the testing phase, the occupied channel of the users and jammer of the last $C$, including the current time slot  are fed of the trained network, and a vector with five elements is outputted from the network. The predicted class of the jammer is 
an index of the outputted vector that has the highest probability value.
\begin{table}[t!]
\centering
\caption{The confusion matrix of two users.}
\label{table:1}
\begin{tabular}[th]{|c|c|c|c|c|c|c|}
\hline
\multicolumn{6}{ |c|  }{Jammer's policy switching time equal to  100}  \\
\hline
 & SJ & RJ & FRJ & RJWD&CJ\\
\hline
SJ&100 & 0 & 0 & 0  & 0 \\ \hline
RJ&0& 100 & 0 & 0  & 0  \\ \hline
FRJ&0& 4 & 96 & 0  & 0  \\ 
\hline
RJWD&0& 1 & 2 & 97  & 0 \\\hline
CJ&5& 0 & 0 & 0  & 95\\ \hline
\end{tabular}
\end{table} 

\section{Simulation Results}

In order to evaluate our proposed jamming recognition method, we simulate the interaction between users and the jammer with considered jamming policies and calculate the accuracy of the detection as a function of the jammer policy switching time. The policy switching time of the jammer is the time during which the jammer maintains a selected jamming policy, and it has been altered from $5$ to $185$ time slots.  We consider a system model with one to four users and twelve available channels. In the considered model, sweeping and combat jammers jam three channels every time slot; also, random and reactive jammers jam  $N$  channels per time slot. In the training process, we consider $T = 5000$, while in the test process, $C$ is assumed to be $20$. Simulation results are averaged over 
$100$ independent runs.

\begin{figure}[t ]
\centering
 \includegraphics[width=0.5\textwidth, height=0.4\textwidth]{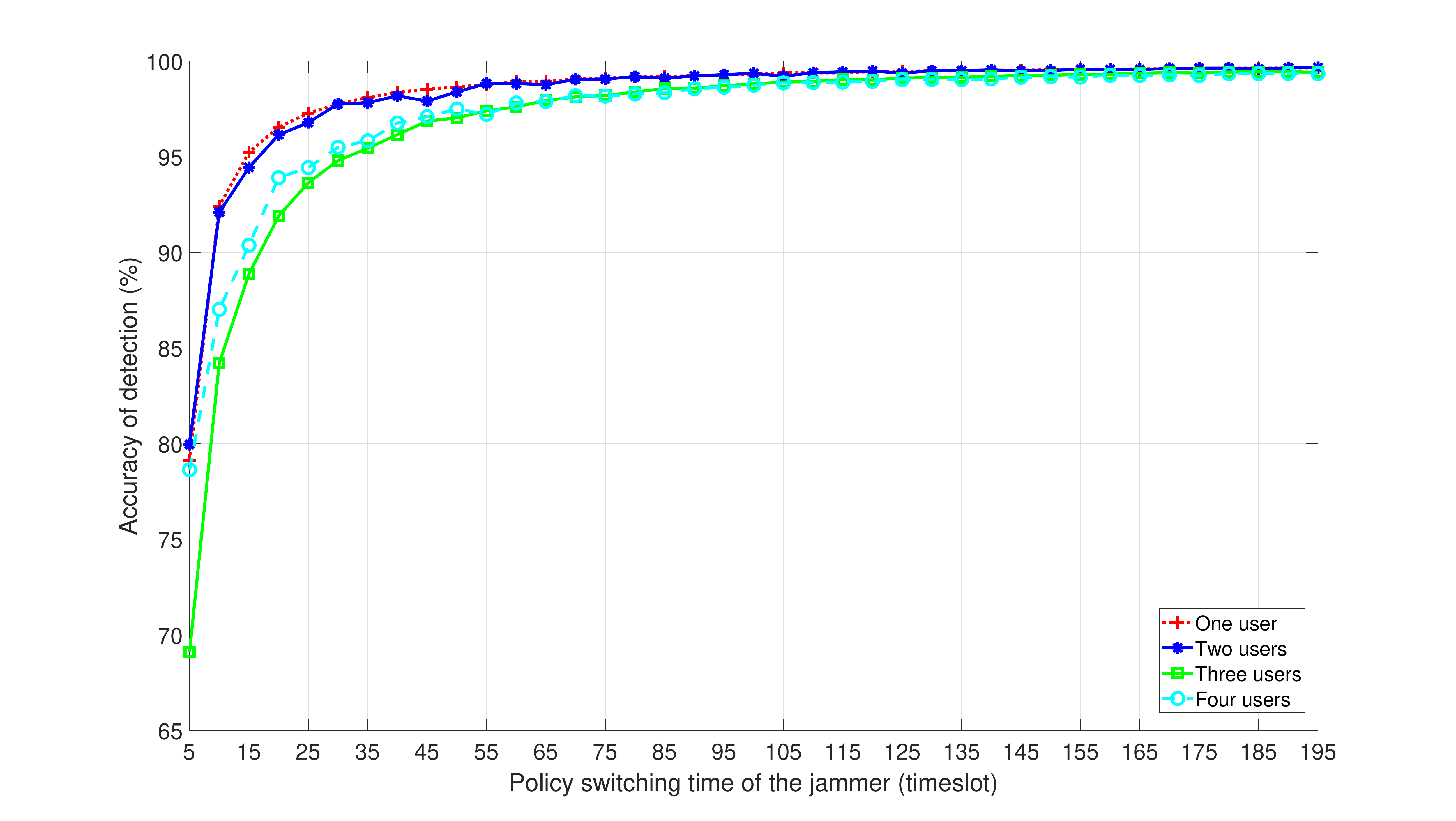}
\caption{The accuracy of the detection as a function of jammer's policy switching time. }
\label{fig4}
  \end{figure}

In Fig. \ref{fig4}, we show the average  accuracy of detection for all the jammer types.  In the figure, when the jammer's policy switch time is five, the detection accuracy for all the considered users' numbers is greater than $70\%$, and when the jammer's policy switch time is $45$, it exceeds $95\%$. As a result, 
the proposed recognition method can detect the jamming policy with high accuracy even when the jammer quickly  changes its policy quickly.
The detection accuracy increases by increasing the jammer's policy switching time for all the considered users' numbers.
This is because, by increasing the jammer's policy switching time, users are provided with more information from each jamming policy to detect the jamming types. The detection accuracy decreases as the number of users increases, except in the case of three users and four users. The reason behind the former is the fact that increasing the number of users increases the detection complexity, and more information should be processed in the trained network. The latter is due to the fact that when the number of users equals three, the number of channels jammed in all jammers' policies is equal.
  
\begin{figure}[t]
\centering
 \includegraphics[width=0.5\textwidth, height=0.4\textwidth]{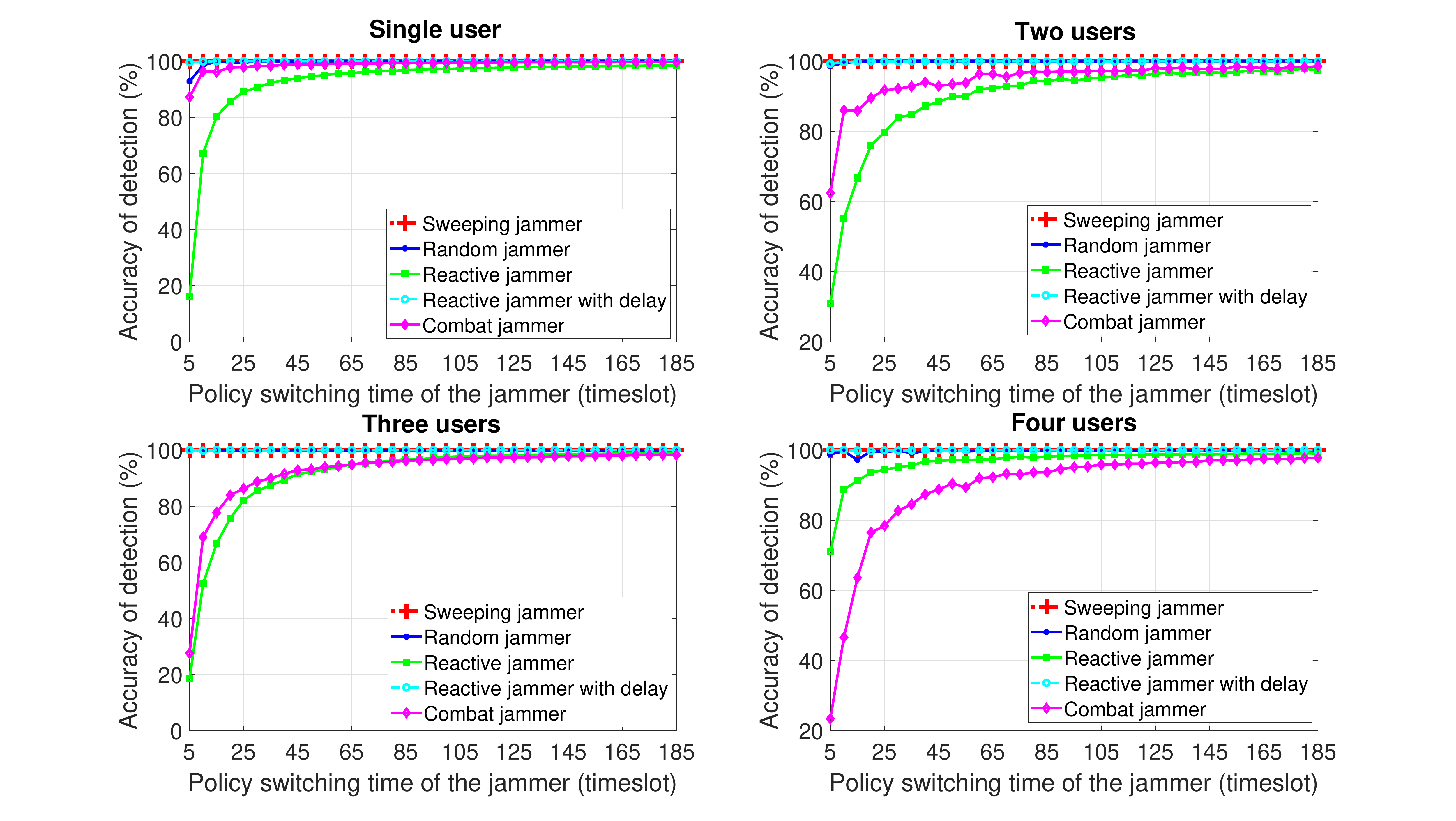}
\caption{Detection accuracy of each jammer type. }
\label{fig5}
  \end{figure}

To clarify the detection accuracy of each jamming strategy under consideration, we exhibit the detection accuracy of each policy independently in Fig. \ref{fig5}. Even though the jammer adjusts its policy every $5$ time slots, Fig. \ref{fig5} demonstrates that sweeping and reactive with delay may be recognized with greater than $95\%$ accuracy. Reactive and combat jamming policies need more time slots, but even in those cases,
 they can be detected with an accuracy higher than $90\%$ 
when the jammer's policy switching time is higher than $45$. 

Misdetection  of the reactive jammer, reactive jammer with delay, and combat jammer can be better justified by the confusion matrix presented in table \ref{table:1}. Table \ref{table:1} shows that, out of $100$ detections, the random jammer is detected instead of reactive jammer four times, fast reactive and random jammers are detected instead of the reactive jammer with delay three times, and the sweeping jammer is detected instead of combat jammer five times, respectively.  The reason behind these confusions is the similarity of the jammer's behavior, especially when the jammer switches from a  policy to another. For instance, the jammer jams three channels per time slot in the considered combat and sweeping jamming policies and  when the jammer changes its jamming policy to combat jammer, the behavior of the jammer becomes very similar to the sweeping jammer.  The same rule holds for reactive and reactive with delay, which causes a miss detection to Random jamming policy.

\section{Conclusion}

In this paper, we have proposed a jamming recognition technique using RNN. To this end, we have considered a general system model, where an AP serves users in the presence of a jammer that changes its jamming policies constantly among five different jamming policies. To recognize the jamming policy, we have proposed  RNN   and outlined the training and testing processes of the proposed RNN. We have simulated the interaction between users and the jammer with considered jamming policies. Simulation results show that all of  the considered jamming policies are detected with high accuracy within a short period in both single-user and multi-user scenarios.  As a result,  the proposed recognition technique can be employed to  immediately detect the jammers' policy, leading to selecting an appropriate anti-jamming method.

  \bibliographystyle{IEEEtran}

\bibliography{ref}
\end{document}